\documentclass{article}

\pdfoutput=1
\usepackage[english]{babel}
\usepackage[utf8]{inputenc}
\usepackage[sort,numbers]{natbib}
\usepackage{hyperref}
\usepackage{graphicx}

\newcommand{\footnoteremember}[2]{
  \footnote{#2}
  \newcounter{#1}
  \setcounter{#1}{\value{footnote}}
}
\newcommand{\footnoterecall}[1]{
  \footnotemark[\value{#1}]
}
 
\title{Conan: a platform for complex network analysis}

\author{Ricardo Honorato-Zimmer\footnoteremember{DLab}{Computational Biology Lab,
Center for Mathematical Modeling, Faculty of Physical and Mathematical Sciences, Universidad de Chile. 
Av. Blanco Encalada 2120, Santiago, Chile.}\footnoteremember{FCV}{Fundaci\'on Ciencia para la Vida.
Avda. Za\~nartu 1482, \~Nu\~noa, Santiago, Chile.} \and Bryan Reynaert\footnoterecall{DLab} \and
Ignacio Vergara\footnoterecall{DLab} \and
Tomas Perez-Acle\footnoterecall{DLab}\footnoterecall{FCV}\footnote{to whom correspondence should be addressed}}

\begin{document}

\maketitle

\begin{abstract}
Conan is a C++ library created for the accurate and efficient modelling,
inference and analysis of complex networks.
It implements the generation and modification of graphs according to several published models,
as well as the unexpensive computation of global and local network properties.
Other features include network inference and community detection.
Furthermore, Conan provides a Python interface to facilitate the use of the
library and its integration in currently existing applications.

Conan is available at \url{http://github.com/rhz/conan/}.
\end{abstract}

\section{Introduction}
The purpose of complex systems science is to understand systems in which a description of
their components is insufficient to explain the system's behaviour.
These kind of systems are found at the centre of several fields of science,
such as physics, biology, and sociology \citep{albert_2002,strogatz_2001}.
Networks are a common representation of those systems, that hold relevant information to study them.
Thus, progress can be achieved in a broad variety of problems relating complex systems throughout
the characterisation, simulation and prediction of the networks structure and dynamics.
Research on this topic is an ongoing effort that has already provided promising results
\citep{demicheli_2010,duarte_2007,borodina_2005}.
However, in spite of the continuous publication of many network-related algorithms and tools,
a unified and modular platform allowing the implementation and use of these theorethical advances is lacking.

There are current applications capable of performing some of the aforementioned tasks,
however, they have two main disadvantages.
First, they are user-oriented, difficulting the integration into high-throughput data analysis workflows.
Second, they are inflexible solutions, not suitable to tackle general problems.
In order to address these shortcomings, we developed Conan, an open-source C++/Python library
which extends the Boost Graph Library to be used as a complex network analysis workbench.
The Boost Graph Library comprises an efficient, well-tested and standard-compliant
set of graph algorithms and data structures.
This allows our library to handle large networks and therefore makes it suitable
for conducting analysis in a wide range of scenarios.
Furthermore, Conan includes many complex network analysis tools in one simple, modular and unified framework.
It spares the scientist the need to use multiple software tools to manage a single problem
and helps them to develop more network algorithms and tools.

\section{Capabilities}
Many algorithms have been developed for the study of complex networks.
The following list summarises those implemented by our library.

\subsection{Network generation}
Conan is able to generate networks based on models published by different authors
\citep{barabasi_1999,watts_1998,demetrius_2005,erdos_1959}, and some canonical models,
such as lattices and rings.

\subsection{Network inference}
Network inference or reverse engineering estimates the interaction networks that underlies observed phenomena.
Several different methods for this purpose have been developed and their performance has been compared \citep{bansal_2007}.
We have implemented some of these in Conan:
\begin{itemize}
  \item Clustering coexpression:
    The covariance of every pair of elements (nodes) in the system is used for selecting the relevant relations.
  \item Maximum entropy (MaxEnt) \citep{lezon_2006}:
    A procedure to identify the pairwise interaction network that has the highest probability of giving rise
    to the observed data.
  \item ARACNe \citep{margolin_2006}:
    The mutual information of each pair of elements in the system is used to infer the network.
    This approach eliminates indirect interactions that are frequently inferred by co-expression methods.
\end{itemize}

\subsection{Topological properties}
\begin{figure}
	\centering\includegraphics[width=0.6\textwidth]{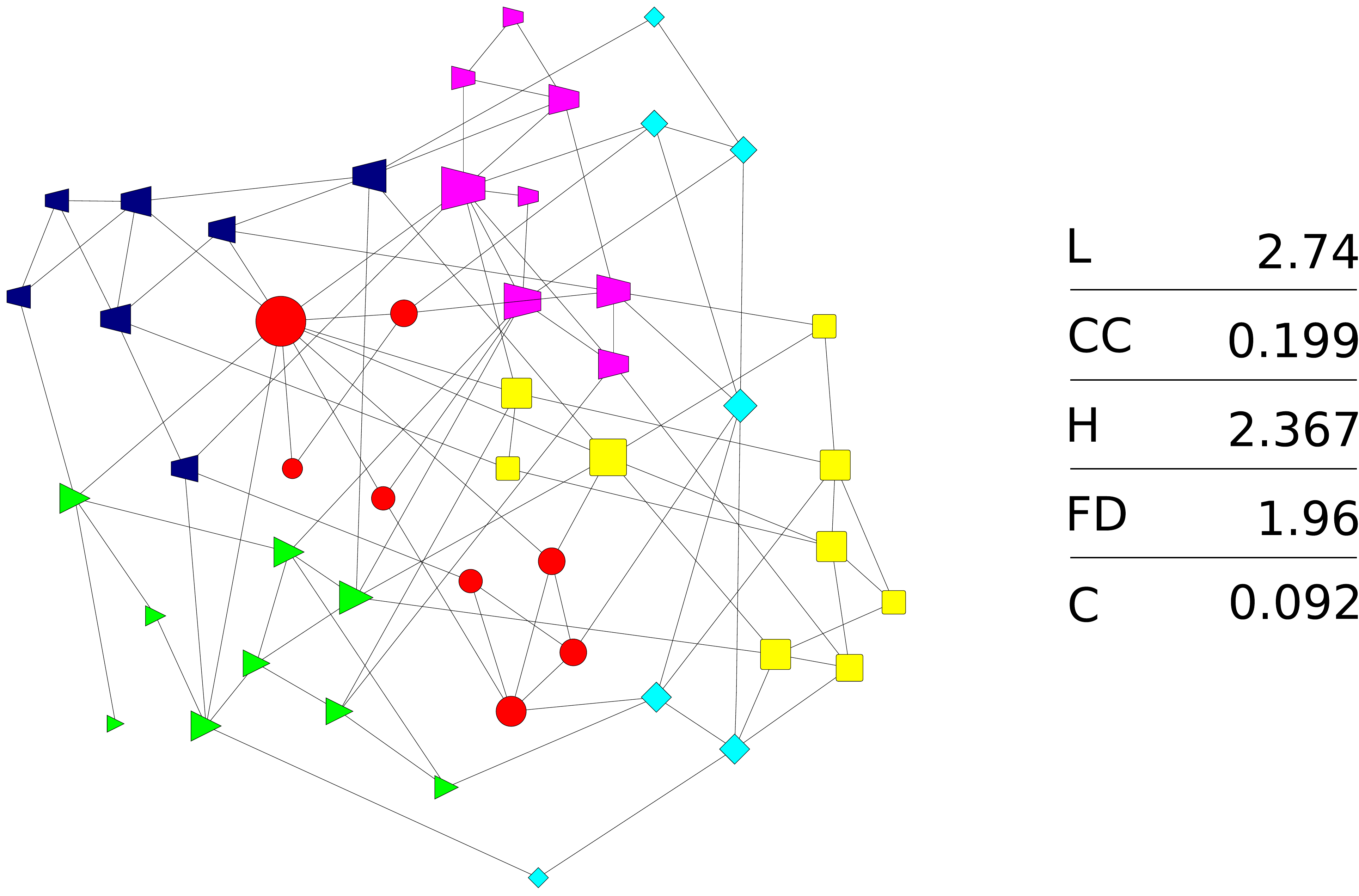}
	\caption{Genetic interaction network of the circadian clock of \textit{Arabidopsis thaliana}
    inferred from microarray data using ARACNe, and some topological properties calculated with Conan.
    Nodes are module-wise coloured and proportional in size to their degree centrality.
    L, average shortest path; CC, average clustering coefficient; H, topological entropy; FD, fractal dimension;
    C, connectance.}
	\label{fig:01}
\end{figure}

Conan can calculate multiple global network properties, such as average shortest path (L),
average clustering coefficient (CC), degree distribution, connectance (C), and many others
(see Figure ~\ref{fig:01}).
The most remarkable are:

\begin{itemize}
  \item Topological entropy \citep{demetrius_2005}:
   This measure quantifies the uncertainty we would have guessing the position of a stochastic walker
   that has been visiting the network for an infinite amount of time.
   It has been correlated to the robustness of the network under random attack. (H in Figure ~\ref{fig:01})
  \item Fractal dimension \citep{song_2005}:
    This quantity, analogous to the euclidean fractal dimension, is an exponent
    which characterizes the self-similar structure of a network.
    It results from the power-law relation that appears under length-scale transformations of the network.
    (FD in Figure ~\ref{fig:01})
\end{itemize}

Conan can also calculate vertex and edge local properties, such as the vertex and edge centrality,
and the vertex clustering coefficient, average shortest path, and eccentricity.

\subsection{Subgraphs and communities}
Conan allows the extraction of user-defined induced subgraphs from a network.
It is also capable of splitting the network into communities using either the method designed by Newman
\citep{newman_2006} or by Guimera and Amaral \citep{guimera_2005}.

\subsection{Multiple format support}
Conan is able to read and write in a variety of file formats, such as GraphML, DOT (\url{http://www.graphviz.org/}),
GML \citep{himsolt_gml}, Pajek \citep{batagelj_1998}, and plain (ASCII) adjacency matrix files.

\section{Interfaces}
Conan is a native C++ library that provides an API for network generation, manipulation and analysis.
The API was designed considering generic programming concepts, which avoids implementation-specific
directives and encourages the development of maximally reusable algorithms.
But, in addition, our library can be utilized from the Python language (\url{http://www.python.org/}).

\subsection{Python Bindings}
    This is the simplest way to access Conan, as it provides an intuitive object-oriented API,
    which lends itself to fast prototyping and simple inclusion within other software packages.
    The bindings' API was written paying close attention to Python's style guidelines.

\section{Acknowledgements}
This work was supported by Fundaci\'on Ciencia para la Vida; and Programa de Financiamiento Basal PFB16.
We are grateful to Hector Urbina for contributing with code snippets.

\bibliographystyle{plainnat}
\bibliography{conan}

\end{document}